# Title: Trading patterns within and between regions – an analysis of Gould-Fernandez brokerage roles

Matthew Smith[1][1] & Yasaman Sarabi[2]

[1]The Business School, Edinburgh Napier University, Edinburgh, UK

[2]Edinburgh Business School, Heriot-Watt University, Edinburgh, UK

**Abstract**

This study examines patterns of regionalisation in the International Trade Network (ITN). The study makes use of Gould – Fernandez brokerage to examine the roles countries play in the ITN linking different regional partitions. An examination of three ITNs is provided for three networks with varying levels of technological content, representing trade in high tech, medium tech, and low-tech goods. Simulated network data, based on multiple approaches including an advanced network model controlling for degree centralisation and clustering patterns, is compared to the observed data to examine whether the roles countries play within and between regions are a result of centralisation and clustering patterns. The findings indicate that the roles countries play between and within regions are indeed a result of centralisation patterns and chiefly clustering patterns; indicating a need to examine the presence of hubs when investigating regionalisation and globalisation patterns in the modern global economy.

1. **Introduction**

In recent years there has been increasing discussions into whether international trade and production patterns organised at the regional or global level (Baldwin and Lopez-Gonzalez, 2018; Emmerij, 1992; Kim and Shin, 2002; Lévy, 2006). Many researchers have made use of sophisticated network analysis techniques to contribute to this debate (De Lombaerde et al., 2018; García-Pérez et al., 2016; Gorgoni et al., 2018a; Tsekeris, 2017; Tzekina et al., 2008). However, many argue that globalisation and regionalisation are not contradictory patterns. Zhu et al. (Zhu et al., 2014) note that the International Trade Network (ITN) is characterised by rich inter and intra-regional training dynamics. Furthermore, regional trading patterns not only contribute to the coordination of regional production networks, they also represent an important stepping stone for countries (or regions) to become further integrated in the wider global economy and increased globalisation patterns (Iapadre and Tajoli, 2014; Slany, 2019). On the other hand, Piccardi and Tajoli (2015, 2012) utilise a community detection approach to the ITN, and do not find significant evidence of regional communities; where they suggest this indicates a truly globalised system. This highlights the need to not only examine the levels of regional and global trade, but to also explore how different regional partitions are linked together.

---

[1] Corresponding Author. E-mail: M.Smith3@napier.ac.uk

Extant literature on regionalisation patterns in the modern global economy is extensive; many examine the emergence of regional supply chains, noting these are often centred on North America, Europe, and Southeast Asia (Pomfret and Sourdin, 2018). Vidya et al. (2020) note that regionalisation is often dominated by developed countries and regions. Southeast Asian regional production patterns have been increasingly examined; where the prominent role of China in the region, and the wider global trading system has given rise to Southeast Asia being at the centre of global regionalisation patterns in the world economy (Andal, 2017; Fan et al., 2019). Athukorala (2011) finds that the rise of global value chains and the fragmentation of the production process has shaped regionalisation patterns in East Asia. He notes that there has been a rise of intra-regional trade of parts and components in East Asia, yet little evidence of growth in intra-regional trade in final goods.

The impact of Regional Trade Agreements (RTAs) on international trade patterns has been examined to inform on regionalisation patterns in the global economy (Reyes et al., 2014; Wu et al., 2020), and RTAs are becoming an increasingly important feature of the modern trading system (Vicard, 2011). Sopranzetti (2018) finds that countries that are more connected to RTAs gain more by increased exports compared to those that are more isolated. However, Garlaschelli et al. (2007) find that trading communities and groups are more likely to arise on the basis of geographic location or market size (GDP) rather than on the basis of RTAs; this finding is also confirmed by Barigozzi et al. (2011) across various commodity specific networks. Others have examined the relationship between geographic distance and trade (Chiarucci et al., 2014; Li et al., 2017; Standaert et al., 2016) and the impact of country borders (Basile et al., 2018), rather than regional partitions or communities. For instance, Abbate et al. (2018) examine patterns of assortativity (the tendency for countries with similar connectivity levels to trade) and clustering (two trade partners of a country, also trade) in the ITN over different distances. They find that the ITN is disassortative over long distance, with highly connected countries linked to peripheral, poorly connected countries, yet it is assortative at short distances. They note that clustering patterns are more evident at shorter distances.

Given the evidence pointing towards patterns of both globalisation and regionalisation, where they are not juxtaposed (Iapadre and Tajoli, 2014; Piccardi and Tajoli, 2015), there is a need to understand the roles that countries play linking both their regions and other geographic partitions together in the ITN. In order to understand the roles played by countries within and between regions, this study draws on Gould – Fernandez brokerage roles (Gould and Fernandez, 1989). This study aims to understand (i) Do the roles that countries play within and between regions significantly emerge beyond degree distribution and clustering patterns? (ii) Do these results differ on the basis of the technological content of traded goods? This study addresses these research questions through the analysis of three international trade networks; trade in high tech, low tech, and medium tech goods, as the link between trading patterns and regionalisation is often sector specific (Ikeda and Watanabe, 2017).

## 2. Data

We extract trade data from UN Comtrade to construct three international trade networks: a high tech, medium tech, and low-tech network for 2018. The definition of the various technological grouping proposed by Lall (2000) was utilised in this study, as the technological content of trade can impact economic growth, development, and productivity opportunities (Aboal et al., 2017). Each trade network is defined as a directed network $G(V, E)$, where $V$ is the set of nodes, which are countries, and $E$ is the set of directed edges, which are the directed trade ties. Trade ties in distinct industries have been found to exhibit substantially different topological characteristics (Cingolani et al., 2018; De Benedictis et al., 2014), therefore in this study high, medium, and low tech networks are examined separately. The backbone of this weighted and directed network is analysed in this paper, where the backbone of the network is extracted using the approach outlined by Serrano et al. (2009). They propose a filtering approach for weighted networks, that retains the significantly heterogeneous links at specified significance level and disregards all other edges. This approach extracts the multiscale backbone as it does not only preserve high value edges, but also preserves low value edges that are important for maintaining the overall connectivity of the network. This backbone filtering approach has been widely applied in the study of international trade networks (García-Pérez et al., 2016; Serrano et al., 2007). In this study, we retain the significantly heterogeneous links at the 0.05 significance level.

Figures (1) to (3) show the visualisation of the trade networks for high, low, and medium tech, where node colour is the regional partition that the country belongs to. The regional partitions are defined according to the World Bank definition, where they provide seven regions: East

Asia & Pacific, Europe & Central Asia, Latin America & Caribbean, Middle East & North Africa, North America, South Asia, and Sub-Saharan Africa. We observe that North American and European nations are at the centre, with countries from Sub-Saharan Africa and (to a lesser extent) Latin America & Caribbean on the periphery.

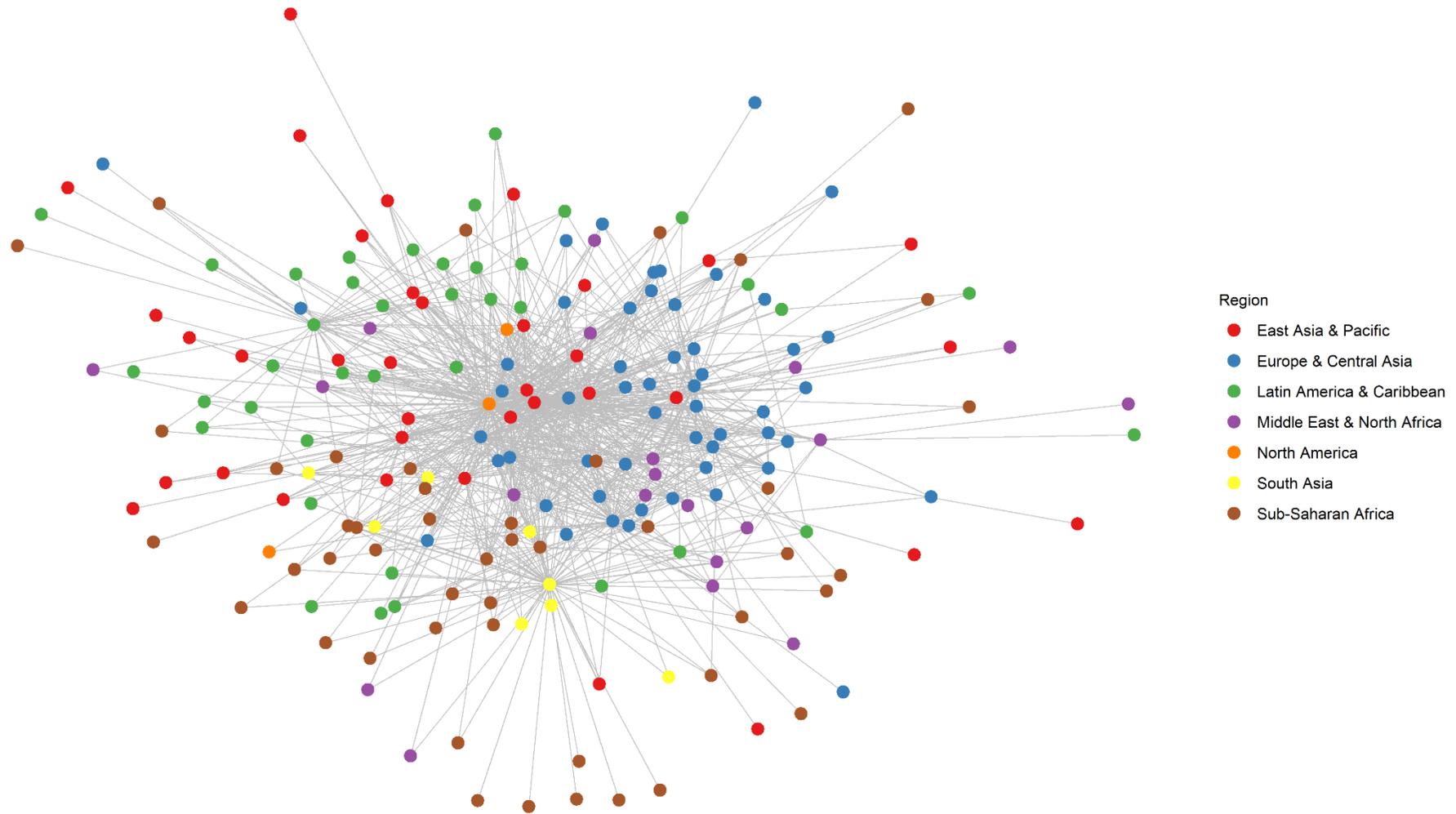

***Figure 1 International Trade Network (ITN) for High Tech***

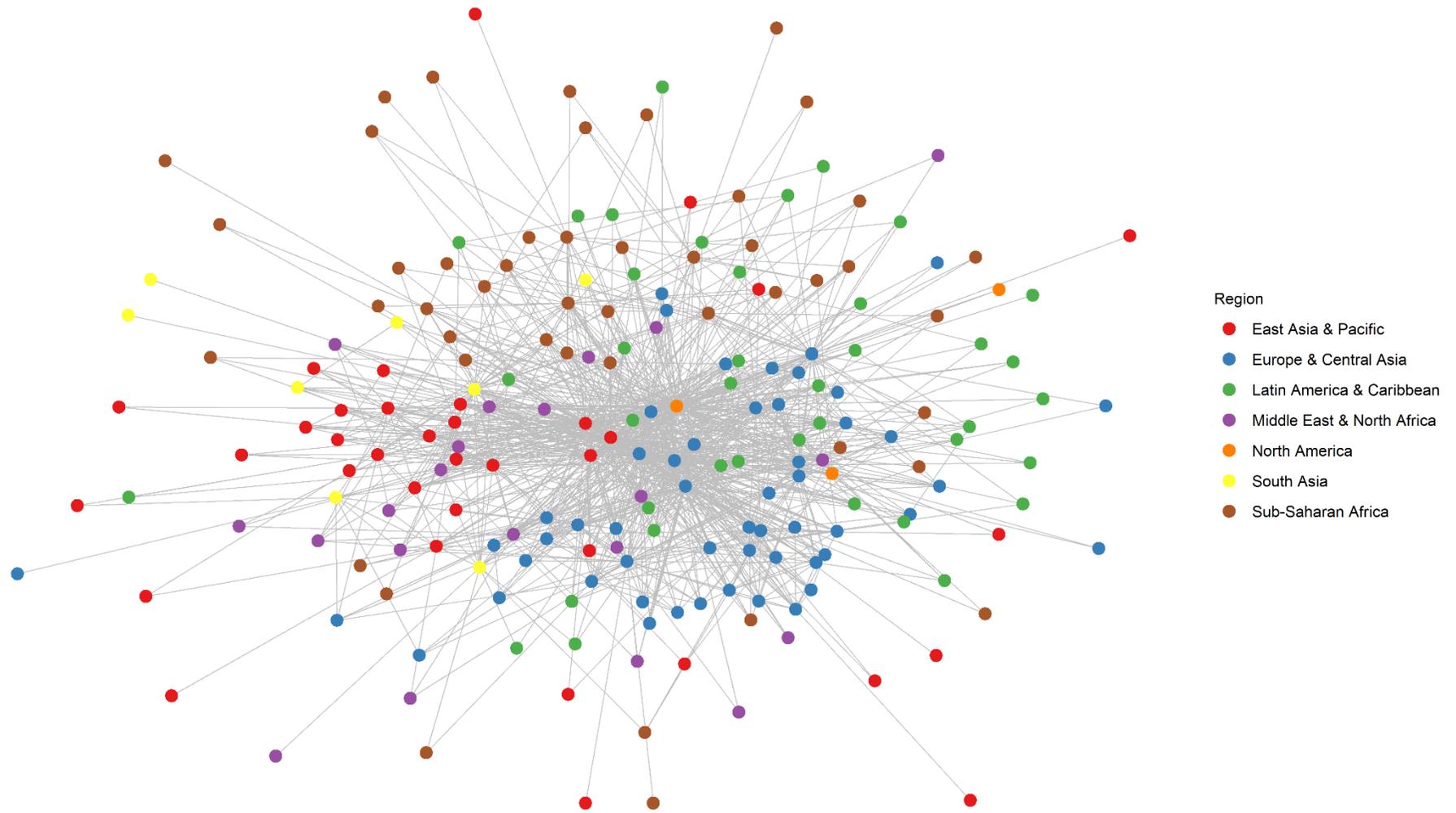

*Figure 2 International Trade Network (ITN) for Medium Tech*

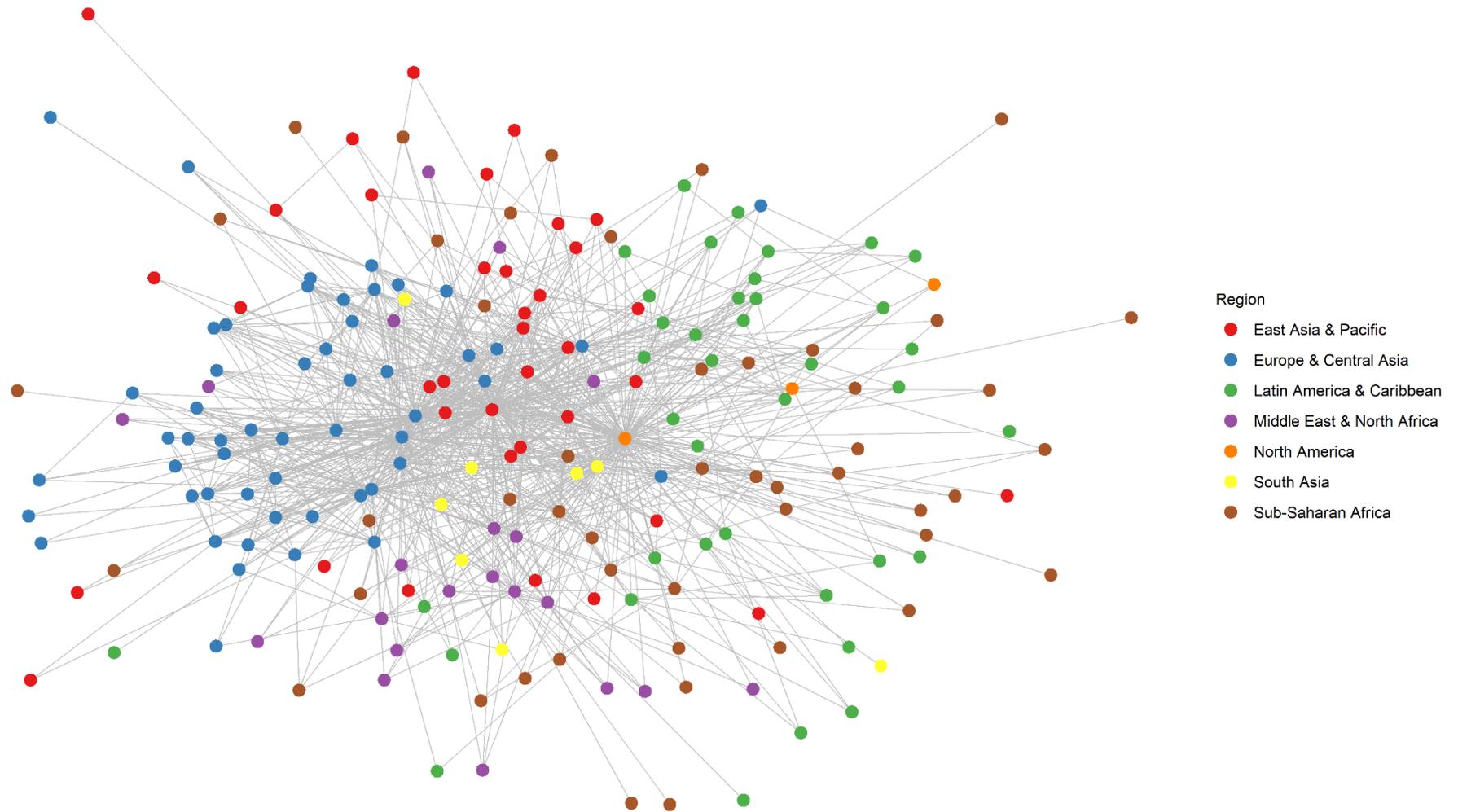

*Figure 3 International Trade Network (ITN) for Low Tech*

## 3. Methods

To address the research questions, we make use of network analysis of the international trade network (ITN). Network analysis is an established methodological approach to understand patterns of international trade (Costa et al., 2011; Fagiolo, 2016; Gorgoni et al., 2018b; Sajedianfard et al., 2021), with numerous empirical applications (Barigozzi et al., 2011; De Benedictis and Tajoli, 2011; de Andrade and Rêgo, 2018; Duan, 2008; Fagiolo et al., 2008; Garlaschelli and Loffredo, 2005; Riccaboni and Schiavo, 2014; Serrano et al., 2007; Serrano and Boguná, 2003; Smith et al., 2016; Smith and Sarabi, 2022; Zhang et al., 2016). To examine the roles countries play within and between regions, addressing the first research question, we make use of Gould-Fernandez (GF) brokerage roles (Gould and Fernandez, 1989); these examine whether an actor plays a specific brokerage role linking actors from its own and different groups. GF brokerage roles are a staple tool in social network analysis, and have been used to tackle research questions in sociology (Stovel and Shaw, 2012), innovation studies (Alberti and Pizzurno, 2015), corporate governance (Sarabi et al., 2021) and economics (Martinus et al., 2021). Amighini and Gorgoni (2014) made use of this approach to better understand the roles countries play within and between regions. Smith and Sarabi (2021) draw on GF roles to examine the role of the UK in the ITN of automotive goods, focusing on how the UK's GF roles may change following Brexit. The GF roles allow for a deeper investigation into regionalisation patterns; for instance, the gatekeeper and representative roles reflect whether inputs are sourced from the same or different regions and whether they are exported to the same or different regions. The consultant role captures whether countries act as an external player to a region. The coordinator and liaison roles reflect regional and global trading patterns respectively. These brokerage roles are presented in Table (1), along with their economic interpretation.

**Table 1 Gould-Fernandez (GF) Brokerage Roles**

| Brokerage Role | Visualisation | Description |
|---|---|---|
| Coordinator | 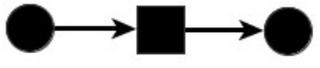 | Coordinators link countries in the same region, deepening regional production sharing. |
| Gatekeeper | 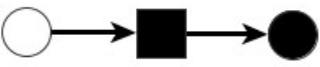 | Gatekeepers import from other regions and then distribute exports in their own region, therefore acting as a regional supplier. |
| Representative | 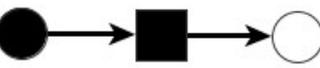 | Representatives import from their own region and export outside the region. These nations act as global distributors for their region. |
| Consultant | 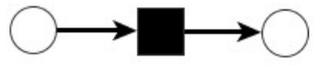 | Consultants link countries from the same region, where they act as external players to regional production networks. |
| Liaison | 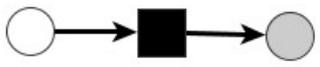 | Countries acting as a Liaison link countries from different regional partitions. |

*Note: The broker is indicated by a square, other nations by a circle. The different colours indicate regional partitions, where the same colour would indicate countries belonging to the same regional partition.*

In order to detect whether these GF role are significant features of the network after controlling for centralisation and clustering, we make use of a variant of the approach developed by Jasny and Lubell (2015) in order to test whether these roles significantly emerge in the international trade network. We first calculate the GF brokerage roles on the observed networks. We then follow the approach outlined by Jasny and Lubell (2015), where we implement an Exponential Random Graph Model (ERGM) (Desmarais and Cranmer, 2012). An ERGM is denoted as (Lusher et al., 2013):

$$P_{\theta,y}(Y = y) = \frac{\exp\{\theta^T g(y)\}}{k(\theta,y)} \quad (1)$$

Where $Y$ is a random variable representing the network, $y$ is the observed network, $g(y)$ is the vector of network terms, $\theta$ is the vector of corresponding coefficients and $k(\theta, y)$ is the normalising constant. The model is estimated with a maximum pseudo-likelihood procedure (for further discussion of this estimation approach see (Desmarais and Cranmer, 2012; van Duijn et al., 2009)). ERGMs have been applied in a wide variety of empirical settings (Chiang, 2015; He et al., 2019; Leifeld, 2018; Traud et al., 2012), including the analysis of international trade and investment patterns (Chu-Shore, 2010; Herman, 2021; Smith et al., 2019).

In the model, we specify structural terms and terms for the tendency of each regional group to export and import. The structural terms include an edges term and a mutual term; the former captures the baseline tendency for ties to form in the network, and the latter the propensity for reciprocated trade ties. A set of centralisation parameters are specified; geometrically weighted in and outdegree centrality parameters. This controls for the in and out degree centralisation in the simulated networks. Further terms include two clustering terms: geometrically weighted dyadwise shared partnership (GWDSP) and geometrically weighted edgewise shared partner (GWESP) (Goodreau, 2007; Hunter, 2007; Snijders et al., 2006). GWDSP captures structural equivalence in the network; the tendency for countries to have identical ties with other countries in the network, yet not necessarily connected to one another. GWESP captures transitivity in the network

In addition to these structural terms, regional sender and receiver terms are included, to control for the tendency for each region to export and import (against the baseline region, Southeast Asia & Pacific). These terms are presented in in Table (2) (please note in the table $x_i$ refers to the regional partition of node *i*). The ERGM is implemented in R using the ergm package (Hunter et al., 2008).

*Table 2 ERGM Terms*

| Term | Statistic |
|---|---|
| Edges | $\sum_{i,j} y_{i,j}$ |
| Mutual | $\sum_{i,j} y_{j,i} y_{i,j}$ |
| Out-Degree (gwodegree) | $\sum_{i=0}^{n} e^{-\alpha y_{i+}}$ |
| In-Degree (gwidegree) | $\sum_{i=0}^{n} e^{-\alpha y_{+i}}$ |
| GWESP | $\sum_{i,j,k} y_{i,j} y_{i,k} y_{j,k}$ |
| GWDSP | $\sum_{i,j,k} y_{i,j} y_{i,k}$ |
| Region Exporter | $\sum_{i,j} x_i y_{ij}$ |
| Region Importer | $\sum_{i,j} x_j y_{ij}$ |

We then use this model to simulate 1000 networks; on each of these networks the GF brokerage roles are calculated for each country, along with the total number of each GF brokerage roles in the simulated networks. This set of simulated networks are regarded as the null model. In order to evaluate the importance of the GF roles in the observed networks, compared to those simulated under the ERGM (controlling for centralisation and clustering), we make use of the Z-Score approach, as proposed by Milo et al. (2004):

$$Z = \frac{f_{real} - \overline{f_{rand}}}{std(\overline{f_{rand}})}$$

The Z-Score is calculated for each GF brokerage role (GF motif), for each ITN, where $f_{real}$ is the number of times the GF role appears in the observed ITN, $\overline{f_{rand}}$ is the average number of times that the GF role appears in the randomised or simulated networks, and $std(\overline{f_{rand}})$ is the standard deviation of the number of times that the GF role appears in the randomised networks. The Z-Score approach is frequently used in network motif discovery (Ahmed et al., 2015).

To examine the impact of degree (against other structural features) and as a check on the ERGM approach, we implement a further alternative approach to simulate a set of trade networks to calculate the GF roles and compare to the observed. In this alternative approach, the null model is created by simulating the network by only preserving the node degree sequence. This allows to examine whether the GF roles are explained by the degree distribution of the ITN. Fagiolo et al (Fagiolo et al., 2013) examine the properties of the international trade network, preserving the node degree sequence. They find that node degree sequences are sufficient to explain a range of higher order network features, including disassortativity and node-clustering correlation. Squartini et al. (2011a, 2011b) in their analysis of the world trade web from 1992 to 2002, found that much of the binary architecture could be explained by a null model controlling for in and out degree. Therefore, we preserve the in and out degree node sequence to examine whether this explains GF roles, reflecting trade patterns within and between regions. These networks are simulated in R using the igraph package (Csardi and Nepusz, 2006). As for the ERGM approach, Z-Scores are also used to evaluate the importance of the GF roles (or motifs) in the observed networks compared to those simulated controlling for the in and out degree node sequence.

4. **Results**

Table (3) contains key network descriptive statistics for each ITN, including density, reciprocity, out degree centralisation, in degree centralisation, and regional assortativity (see Borgatti et al., 2018 for a full definition of these metrics). Regional assortativity captures the correlation of regional partition membership between connected countries in the network. A positive score would point towards regional trade, whilst negative towards global trade.

Centralisation is a network level metric that captures the dispersion of centrality scores in the network (relative to the most central score in the network) (Everett et al., 2004; Sinclair, 2011, 2009).

*Table 3 Descriptive Network Statistics*

| Metric | High tech | Medium Tech | Low Tech |
|---|---|---|---|
| **Size (Number of countries)** | 206 | 207 | 209 |
| **Density** | 0.0269 | 0.0328 | 0.0294 |
| **Reciprocity** | 0.2396 | 0.3043 | 0.3088 |
| **In Degree Centralisation** | 0.3048 | 0.3021 | 0.4466 |
| **Out Degree Centralisation** | 0.5048 | 0.4672 | 0.5043 |
| **Regional Assortativity** | 0.2283 | 0.3064 | 0.3516 |

We observe consistent patterns across the trade network with varying technological content. We observe a higher out degree centralisation than indegree centralisation; indicating that export ties are more likely to be concentrated in a handful of key countries than import ties. Regional assortativity is positive across the three networks, pointing towards regional (rather than global) trade.

Figures (1) to (3) indicate consistent patterns across the three ITNs; they have a core-periphery structure, with a set of countries tightly connected at the centre, with weakly connected nations on the periphery (Borgatti and Everett, 2000). To examine this in further detail, we employ the core-periphery algorithm suitable for weighted and directed network developed by Ma and Mondragón (2015), in order to categorise a country as a member of the core or periphery in these three international trade networks. This approach ranks nodes based on decreasing order of node strength (the weighted degree centrality), $\sigma_i$. A node with rank $r$ and node strength $\sigma$ is referred to as $\sigma_r$. The quantity of the node strength arising from linking to higher ranked nodes is calculated and referred to as $\sigma_r^+$. There will be a node $r^*$ where $\sigma_r^+$ has reached its maximum, and from that node onwards, $\sigma_r^+$ will always be less than $\sigma_{r^*}^+$ ($\sigma_{r^*}^+ > \sigma_r^+$). This is used to identify the boundary between the core and the periphery, where all nodes with a rank less than or equal to $r^*$ are included in the core and all other nodes categorised as belonging to the periphery. Figure (4) presents three world maps, indicating whether a country is a member of the core or periphery for the three networks. We observe a smaller core for the case of high-tech trade, yet this is larger and consists of more European nations for medium and low tech.

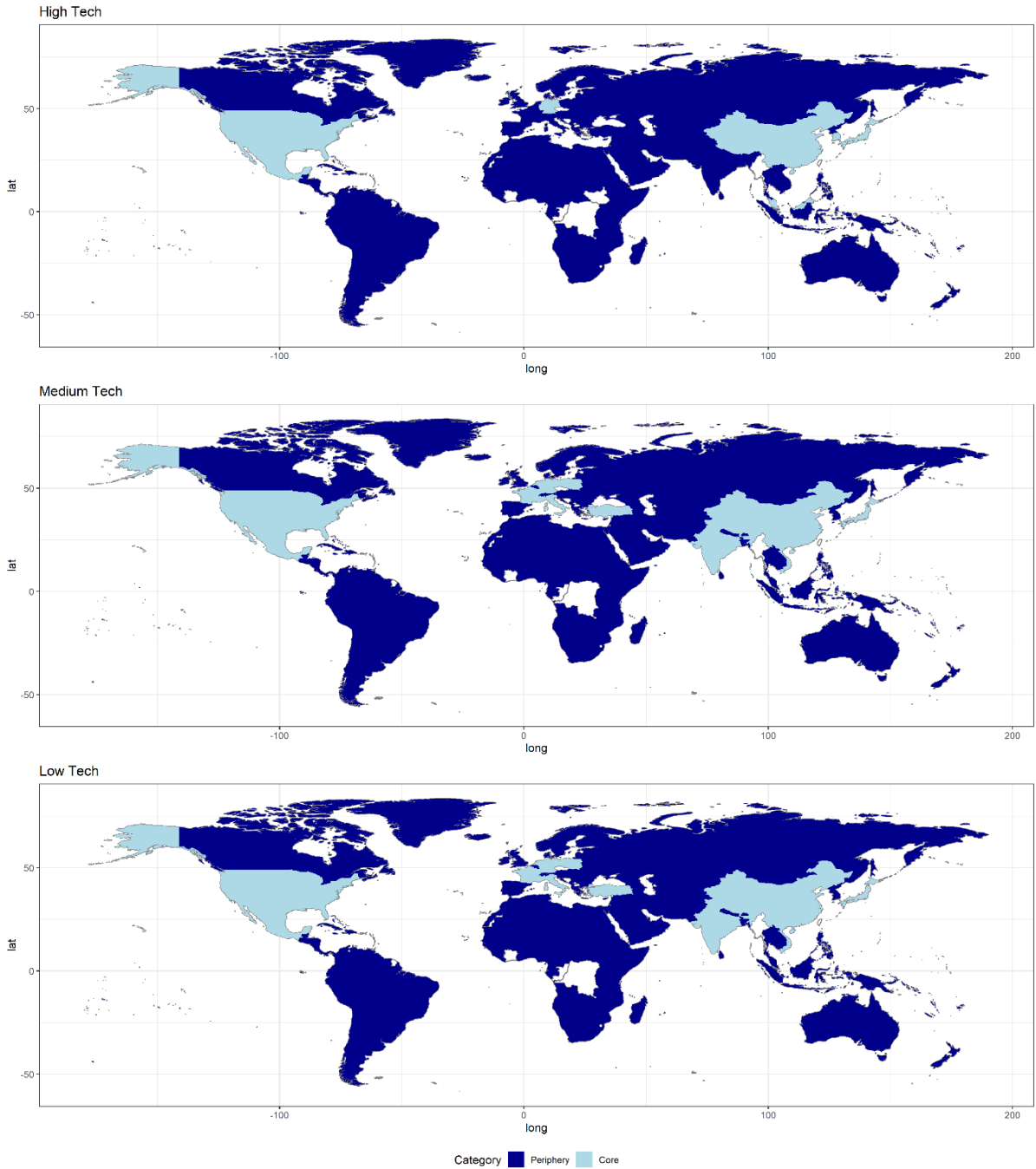

*Figure 4 Core-Periphery Analysis*

The External-Internal (E-I) Index (Krackhardt and Stern, 1988) propose a measure to capture whether an actor's ties are more internal or external to their group membership. This has been utilised for international trade (Amighini and Gorgoni, 2014; Smith et al., 2016) and FDI (Bolívar et al., 2019) networks to examine patterns of regionalisation, where group membership is regional partition.

The E-I Index is defined as:

$$E - I\ Index = \frac{E-I}{E+I}$$

Where $E$ is the number of ties an actor has external to the group, and $I$ is the number of ties an actor has internal to the group; values can range from 1 to -1. In the case of the ITN, where group membership is the regional partition, a positive value would indicate the country tends to trade outside the regional partition, and a negative value would suggest the country has a greater number of regional trade ties. Figure (5) provides a map of the country E-I index results for the three ITNs; positive E-I scores are visualised in red, and negative E-I scores are blue. We observe that negative E-I scores, represented by blue countries tend to be concentrated in Europe for all three component groups. Figure (5) indicates that parts of South East Asia & Pacific have a negative E-I index score, which reflects a level of regionalisation, however China consistently has a positive E-I index, reflecting its role as a more global player in these networks (and the region). Differences between the component groups can be observed in South America and Sub-Sahara Africa. In South America, there is a tendency for more global (compared to regional) trade in high tech goods (compared to medium and low tech). Similar patterns are also observed in Sub-Saharan Africa.

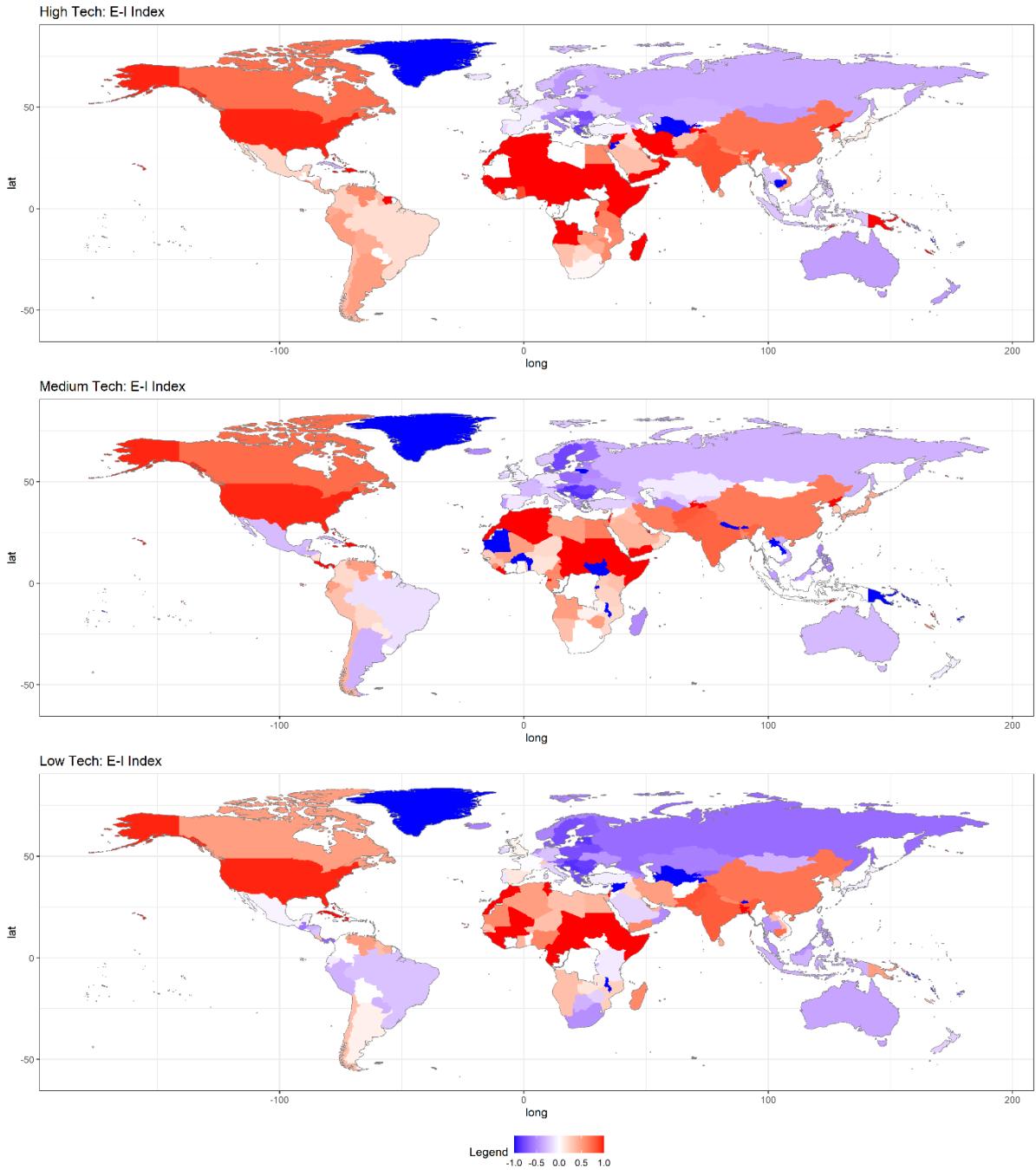

*Figure 5 E-I Index Results*

The GF roles were calculated for the three observed international trade networks. Figures (6) to (8) compare the GF brokerage roles with the (total) degree centrality of the country for each international trade network. We can observe that there are a number of periphery countries in the three ITNs that have both low degree and low GF brokerage roles.

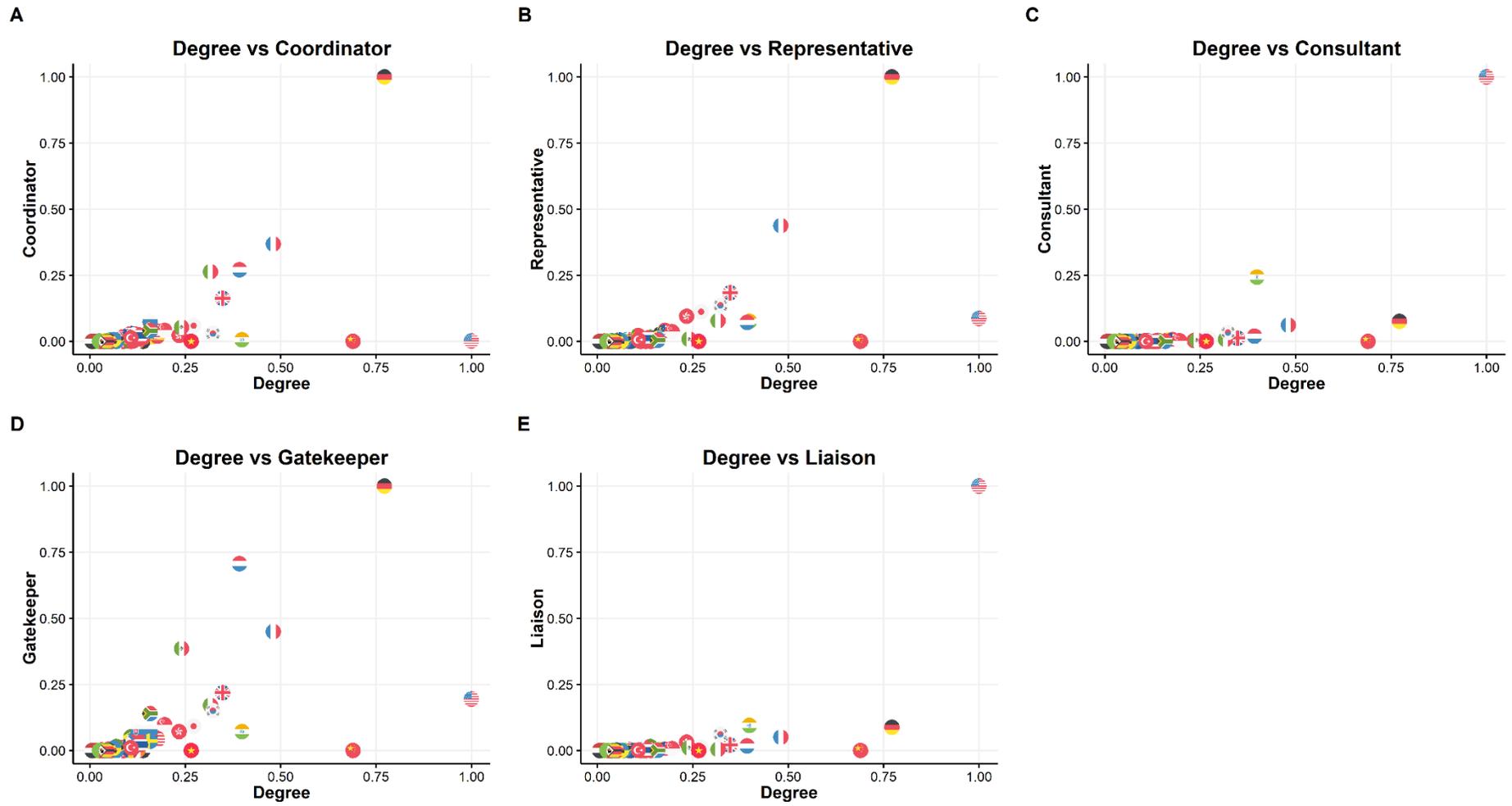

*Figure 6 Degree vs GF Roles - High Tech*

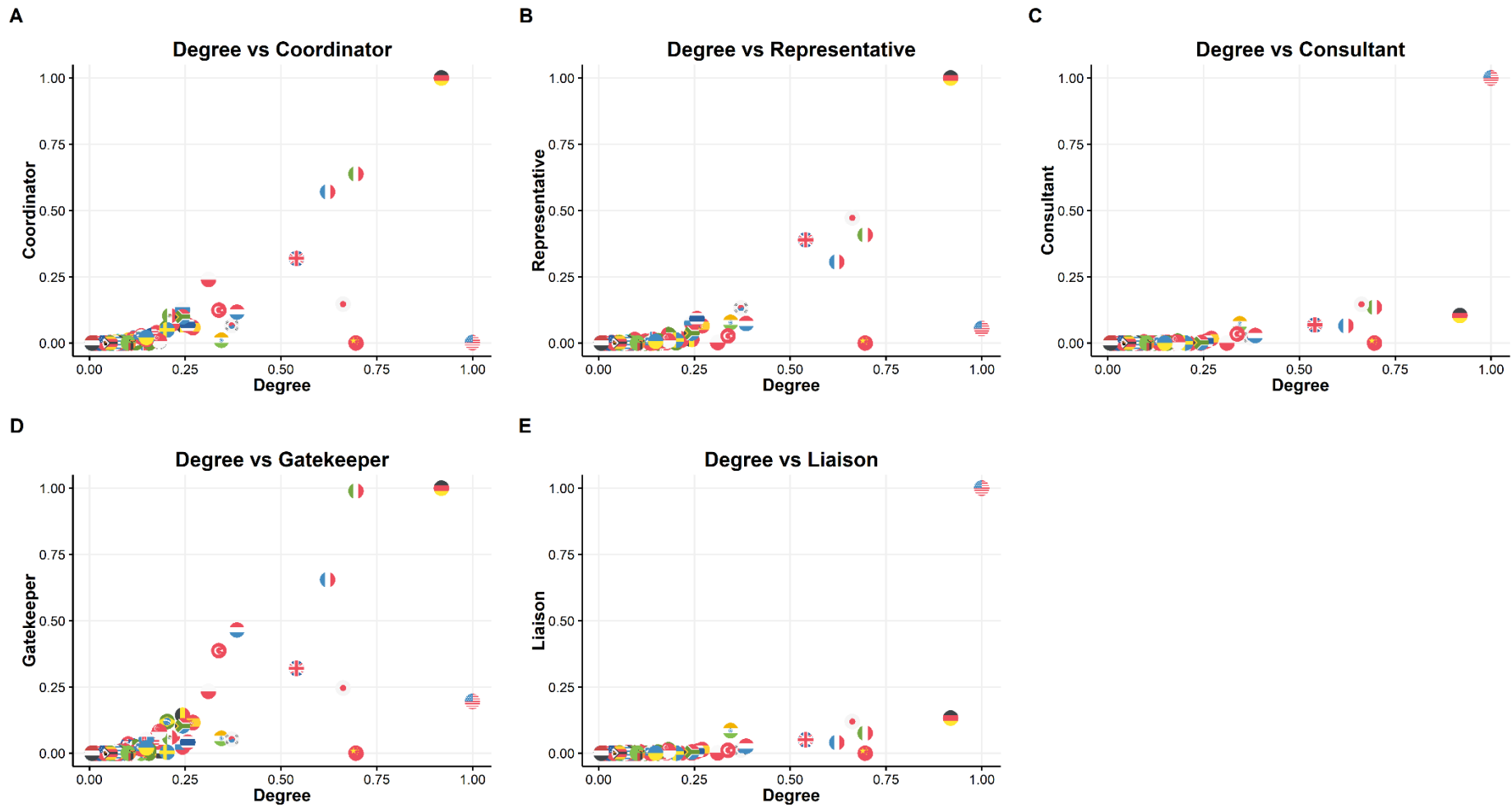

*Figure 7 Degree vs GF Roles - Medium Tech*

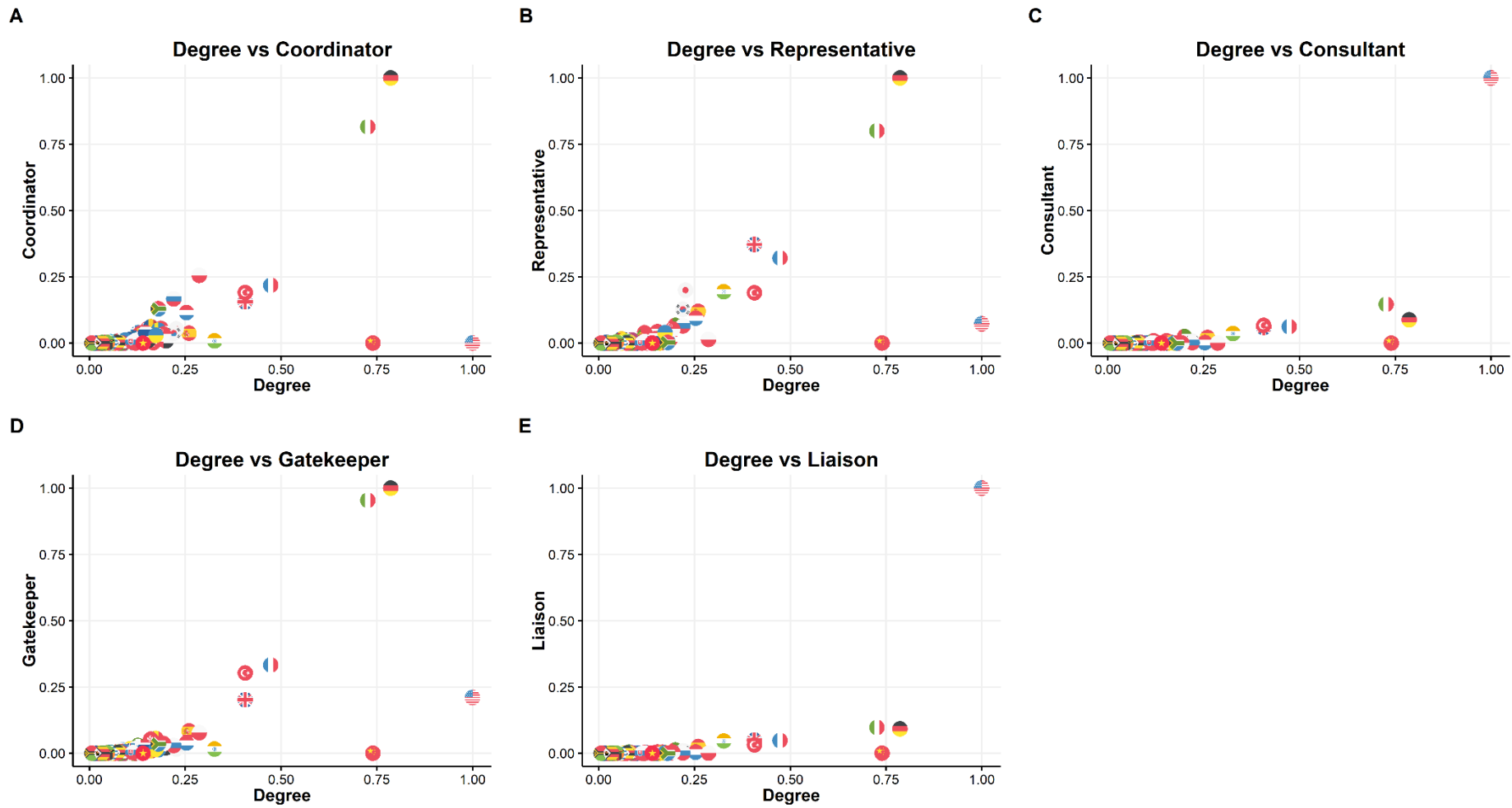

*Figure 8 Degree vs GF Roles - Low Tech*

The patterns appear to be consistent across the three component groups, where the coordinator, gatekeeper and representative roles (which are associated with regional trade) are more closely related to degree scores, whilst liaison and consultant (which are more associated with global trading patterns) are less so.

The ERGM is then implemented in order to construct the simulated networks. The results from the ERGM are presented in Table (4), where the estimated coefficients are given with the standard error in parenthesis. The baseline region for the regional export and importer effects is Southeast Asia & Pacific.

Only a brief overview of the ERGM results is presented here, as the ERGM is chiefly used for network simulation, where additional variables would be required to properly model trade tie formation. The structural parameters included in the models are consistent across the three trade networks. The mutual parameter is positive and significant, indicating that these networks are highly symmetric. Both the out and in degree centralisation parameters are negative and significant, this indicates that the network has a centralised structure, with both import and export ties concentrated in a handful of countries in the network (in line with findings from extant literature) (Li et al., 2003). This pattern is more pronounced for imports, as reflected by the gwidegree result. This suggests that there are hubs in the trade network, which may subsequently shape the roles that countries play within and between regions. The results indicate that there are hubs in all three networks, regardless of the technological content of the products being traded.

The clustering effects (GWESP and GWDSP) differ somewhat between the different networks; there is a positive and significant GWESP parameter across groups, yet varying GWDSP results, ranging from positive and significant in the high-tech group, negative and significant in medium, and non-significant in the low-tech grouping. The GWESP results point towards clustering tendencies across the product groups. A positive GWESP and negative GWDSP would indicate that there is a tendency for triadic closure in the network, where open triads tend to close. We observe this in the medium tech grouping, which suggests that countries that share trade patterns are also likely to trade. In the case of the high-tech network, the GWESP indicates that there is a tendency for clustering, yet there is a tendency for open triads, where countries sharing trade partners, are not forming a trade tie between them.

The significant centralisation and clustering effects highlight the need to account for this when examining brokerage roles, and whether regional or globalisation trading patterns emerge above and beyond these structural patterns.

*Table 4 ERGM Results*

|  | High-Tech | Medium-Tech | Low-Tech |
|---|---|---|---|
| Edges | -3.2167*** | -4.1816*** | -4.2532*** |
|  | (0.1241) | (0.1343) | (0.1294) |
| Mutual | 1.5963*** | 1.9195*** | 1.7956*** |
|  | (0.1105) | (0.0948) | (0.0981) |
| Out-Degree (gwodegree) | -2.0472*** | -1.1373*** | -0.7292*** |
|  | (0.2191) | (0.2329) | (0.2180) |
| In-Degree (gwidegree) | -5.8121*** | -5.3605*** | -4.7786*** |
|  | (0.5791) | (0.7249) | (0.5576) |
| GWESP | 0.7266*** | 1.3820*** | 1.2204*** |
|  | (0.0439) | (0.0528) | (0.0457) |
| GWDSP | 0.0178*** | -0.0056* | 0.0029 |
|  | (0.0030) | (0.0027) | (0.0031) |
| Europe & Central Asia Exporter | -0.4667*** | -0.4867*** | -0.2417** |
|  | (0.0855) | (0.0828) | (0.0878) |
| Latin America & Caribbean Exporter | -1.1719*** | -1.2909*** | -0.9476*** |
|  | (0.1306) | (0.1213) | (0.1283) |
| Middle East & North Africa Exporter | -0.9831*** | -1.3710*** | -1.0242*** |
|  | (0.1527) | (0.1425) | (0.1422) |
| North America Exporter | 0.5414** | 0.2911 | -0.8841** |
|  | (0.1787) | (0.1899) | (0.2934) |
| South Asia Exporter | -0.9903*** | -1.0410*** | 0.1066 |
|  | (0.1977) | (0.2154) | (0.1499) |
| Sub-Saharan Africa Exporter | -1.1799*** | -1.1446*** | -1.0791*** |
|  | (0.1320) | (0.1184) | (0.1211) |
| Europe & Central Asia Importer | 0.0354 | 0.1990 | 0.2151* |
|  | (0.1006) | (0.1050) | (0.1068) |
| Latin America & Caribbean Importer | -0.1003 | 0.1915 | 0.0043 |
|  | (0.1336) | (0.1276) | (0.1333) |
| Middle East & North Africa Importer | -0.0242 | 0.5572*** | 0.3531* |
|  | (0.1605) | (0.1427) | (0.1463) |
| North America Importer | -0.1014 | 0.6435** | 0.9744*** |
|  | (0.2255) | (0.1974) | (0.1724) |
| South Asia Importer | 0.8476*** | 0.6974*** | -0.6616* |
|  | (0.1673) | (0.1796) | (0.2592) |
| Sub-Saharan Africa Importer | 0.1709 | 0.5163*** | 0.2707 |
|  | (0.1333) | (0.1285) | (0.1406) |
| AIC | 6791.8224 | 7433.4569 | 7079.1686 |
| Log Likelihood | -3377.9112 | -3698.7285 | -3521.5843 |

***p < 0.001, **p < 0.01, *p < 0.05

The exporter and importer results indicate whether countries from the regions are more or less likely to form trade ties than Southeast Asia & Pacific, the baseline region. Table (4) indicates that regions tend to export significantly less than Southeast Asia & Pacific in all product group with the exception of North America, and to a lesser extent South Asia. For North America, there is only a weakly significant exporter effect in the high-tech group, where it is negative and significant in low-tech. This indicates that for low tech goods, countries from North America are less likely to export these products than nations in Southeast Asia & Pacific. In the case of South Asia, there is no significant difference in low tech category; perhaps indicating the region has a comparative advantage in the production of these goods.

There appears to be more variation between product groups and regions in terms of import patterns, as observed in Table (4). In the high-tech group, only South Asia imports significantly more than Southeast Asia & Pacific, whilst the other regional import patterns are not significantly different. This is a clear contrast to the products with medium and low technological content. In the medium tech grouping, Middle East & North Africa, North America, South Asia, and Sub-Saharan Africa each import significantly more than Southeast Asia & Pacific. For low tech, there is a weakly significant European importer effect, indicating that only in this product group does Europe import greater levels, indicating a tendency for European nations to import and source low tech products. These patterns are also observed for North America (where this pattern is more pronounced) and the Middle East & North Africa. South Asia imports significantly less (at the 0.05 level) than Southeast Asia & Pacific, reflecting the strong supplier chain and comparative advantage of low-tech products in the region, where countries here are less likely to source low tech goods (Ando, 2006; Mayer and Wood, 2001).

The ERGM models are then used to simulate 1000 networks for each product group. The GF roles are calculated for each of these 1000 networks. The Z-score is then calculated for each GF role for each component grouping. Figure (9) presents the Z-Score results, presenting the significance profile for each GF role for each component group. When $Z < -2$, which is for the majority of GF roles, this indicates that the GF roles occur less frequently in the observed network than the simulated networks; Ohnishi et al. (2010) would refer to these GF roles as "anti-motifs". These GF roles are "anti-motifs" for all three component groups; the one exception is the liaison role for the low-tech group. This indicates that many global linkages are results of centralisation and clustering patterns. This suggests that there are key hubs playing an important role in globally connecting world trade network. This also indicates that when measures and metrics are utilised to map the globalisation trends in world trade, there is a need to acknowledge and account for the trading patterns of a few major player, where high levels of trade are concentrated. These results are in line with the work of Abbate et al. (2018), in which they find that hubs of the network are usually connected to countries far apart. This suggests that centralisation and clustering tendencies explain the rise of coordinators across product groups, and points towards being a global hub contributing to a strong regional position; integrating in the network at both the global and regional level.

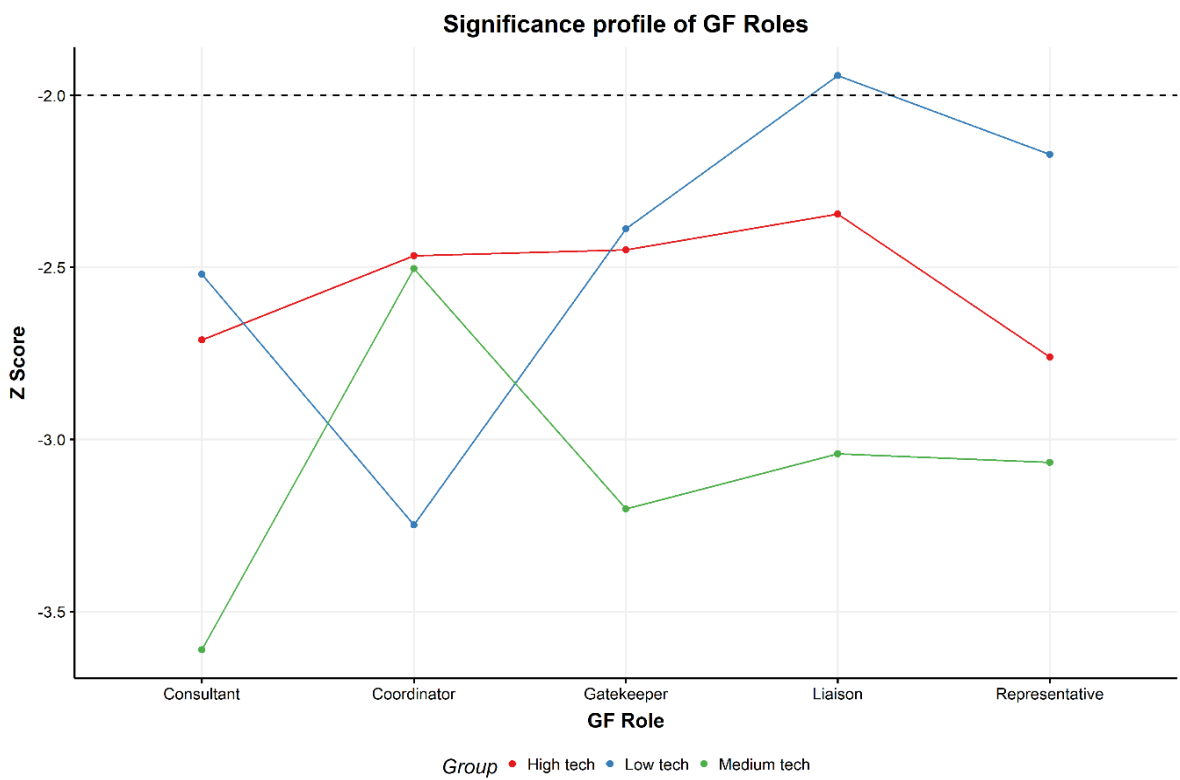

*Figure 9 Z-Scores for GF Roles - ERGM Null Model*

In order to examine whether the roles are a result of the degree of a country, we implement an alternative approach, where the networks are simulated fixing the node degree sequence: Figure (10) provides the Z-Score results.

The results contrast to those presented in figure (9), using the ERGM simulations as the null model. With the exception of the consultant role for medium and low tech, $Z > 2$, this indicates that the GF roles occur more frequently in the observed network than the simulated networks. This suggests that clustering patterns (along with other structural features), rather than the degree sequence alone, contributes to explaining the roles that countries play within and between regions. The Z scores also follow a more consistent pattern across component groups compared to the results presented in figure (9). The consultant role is non-significant in the case of consultant roles for medium and low tech, this suggests the role a country plays linking to countries from another region is a result of degree patterns. Yet the coordinator role, which captures whether a country is integrated into the region, is not a result of degree patterns alone.

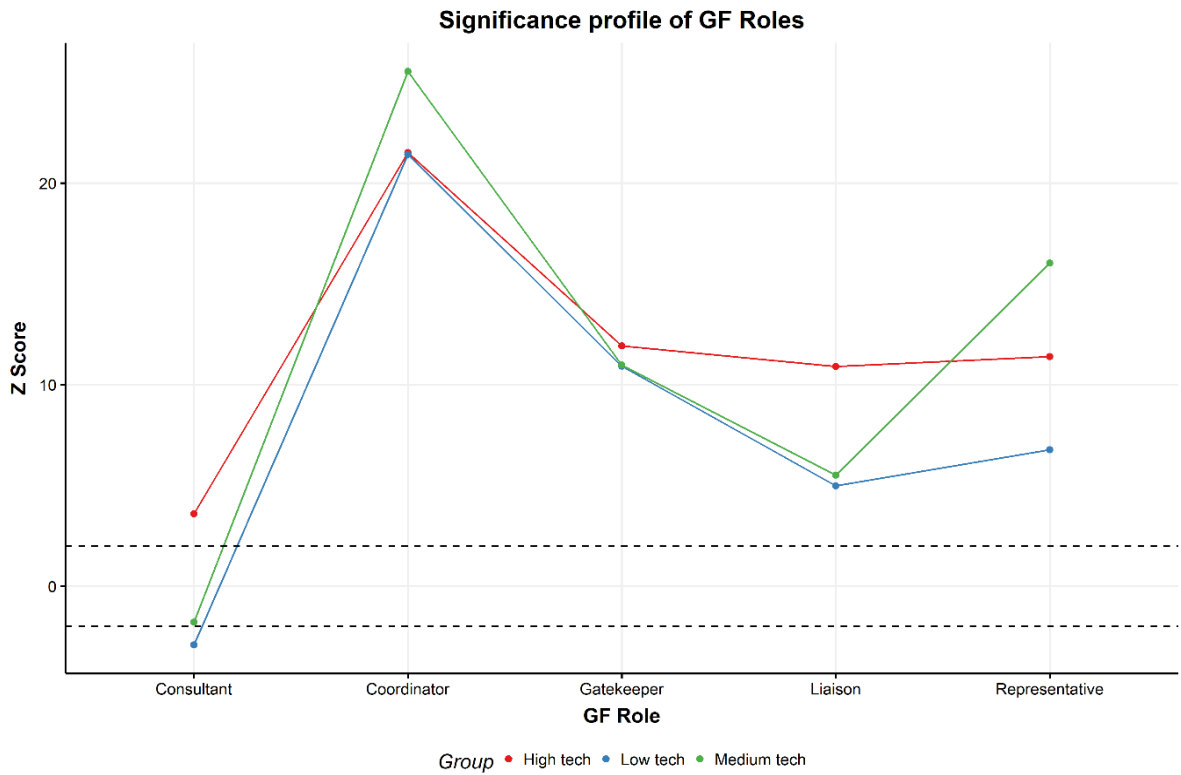

*Figure 10 Z-Scores for GF Roles - Fixed Degree Sequence Null Model*

These findings suggest that there is a need to account for clustering patterns, (and to a lesser extent degree centrality) when developing indices to measure regionalisation (or globalisation) patterns in the modern world economy.

## 5. Discussion & Conclusion

This study aimed to examine the roles that countries play within and between regions and whether they significantly emerge beyond degree distribution and clustering patterns. This study further aimed to examine whether these patterns differ on the basis of the technological content of traded goods. This study addressed these research questions through the analysis of three international trade networks: trade in high tech, low tech, and medium tech goods.

The results from the GF analysis indicate that the roles that countries play within and between regions is a result of centralisation and clustering tendencies, regardless of the technological content of trade flows. The results suggest that patterns of regional trade, as indicated by the GF roles, are not countries playing unique roles in the network and their region, rather are results of the major trade centres in the modern global economy, such as United States, China, Japan, and many key European nations (Deguchi et al., 2014; Fan et al., 2014; Ferrarini, 2013). These findings are in line with the work of Bao and Wang (2019); they note that the self-strengthening effect of hubs in the trade network has positive country level effects, and has led to the global expansion of RTAs.

One limitation of this work is that it is restricted to trade based on technological content. Future analysis could be applied to various other types of trade, such as intra-industry or inter-industry trade. Alternatively, future work could apply the analysis to networks of value-added to understand whether the interplay between the roles countries play between and within regions, and patterns of node degree, centralisation and clustering differ in various economic networks.

This study highlights a need to better understand patterns of regionalisation in the global economy, given that many policy debates centre on the levels of integration at the macro-regional level (as demonstrated by Brexit) (Frigant and Zumpe, 2017). It further highlights that if GF roles are utilised to understand trading patterns, (as observed in Amighini and Gorgoni, 2014; Gorgoni et al., 2018a) there is a need to acknowledge the role of hubs and major trading centres in contributing to the emergence of these roles. A potential policy recommendation is that countries, especially developing countries, should strengthen hubness, which could allow a country to reshape the impact from regionalisation and globalisation patterns.